\documentclass[fleqn,twoside]{article} 
\usepackage{espcrc2} 
\usepackage{epsfig}
 
\usepackage{graphicx} 
\usepackage[figuresright]{rotating} 
 
\newcommand{\AmS}{{\protect\the\textfont2 
  A\kern-.1667em\lower.5ex\hbox{M}\kern-.125emS}} 
 
\title{Chirally improved Dirac operators: Studying the sensitivity to  
topological excitations for zero and finite temperature\thanks{Based on
the contributions of C.~Gattringer, C.B.~Lang and S.~Schaefer at Lattice 2001.}} 
 
\author{Christof Gattringer\address[Regensburg]{Institut  
f\"ur Theoretische Physik, 
Universit\"at Regensburg, 
D-93040 Regensburg, Germany}\thanks{Supported by the Austrian  
Academy of Sciences (Apart 654).}, 
Meinulf G\"ockeler\addressmark[Regensburg], 
C.B.~Lang\address[Graz]{Institut f\"ur Theoretische Physik,
Karl-Franzens-Universit\"at Graz, 
A-8010 Graz, Austria}, 
P.E.L.~Rakow\addressmark[Regensburg], 
Stefan Schaefer\addressmark[Regensburg] \\and 
Andreas Sch\"afer\addressmark[Regensburg]}

\newcommand{\be}{\begin{equation}} 
\newcommand{\ee}{\end{equation}} 
\newcommand{\sign}{\mathop{\mathrm{sign}}\nolimits}

\begin{document} 
 
\begin{abstract} 
We discuss the construction and properties of an approximate solution of the
Ginsparg-Wilson equation, the so-called chirally improved lattice  Dirac
operator. In particular we study the behavior of its eigenmodes in smooth
instanton backgrounds as well as for thermalized gauge configurations on both
sides of the QCD phase transition. We compare with results from other Dirac
operators including the overlap operator.  The results support the picture of
chiral symmetry breaking being closely related to instantons.
\vspace{1pc} 
\end{abstract} 
 
\maketitle 
\section{Motivation and introduction} 
 
During the last three years it has been realized that the Ginsparg-Wilson 
relation \cite{GiWi82} is the key to chirally symmetric fermions on the
lattice.  However, the overlap operator \cite{overlap} which exactly solves the
Ginsparg-Wilson equation is numerically very expensive. For many applications
an alternative to an exact solution is an approximate Ginsparg-Wilson Dirac
operator, such as a finite parametrization of the perfect action
\cite{perfect}  or the recently proposed chirally improved Dirac operator
\cite{chirimpact}.  The latter results from a systematic expansion of a
solution  of the Ginsparg-Wilson equation. 

In this contribution we report on our results for the eigenvectors and
eigenvalues of the chirally improved Dirac operator, using smooth instanton
backgrounds as well as thermalized (quenched) gauge configurations on both
sides of the QCD phase transition. These studies serve two purposes: Firstly,
they allow to obtain a sound understanding of the properties of the chirally
improved Dirac operator. In particular we also compare our results to
calculations done with  the overlap operator in order to understand the effects
of small violations of chirality. Secondly, the analysis of the eigenmodes
provides insights into the nature of the QCD vacuum, since the eigenvectors
couple to localized objects in the underlying gauge field such as instantons or
instanton anti-instanton  molecules.

\section{Remarks on the construction of the lattice Dirac operators}

{\bf Chirally improved operator:} 
In \cite{chirimpact} we suggested an expansion of the lattice  Dirac operator
$D$ in terms of a systematic series of simple hopping terms,  connecting sites
along paths of arbitrary length and allowing for all  elements of the Clifford
algebra.  For the ansatz one respects all standard translational and rotational
symmetries   of the lattice Dirac operator as well as invariance under C and P
and  $\gamma_5$-hermiticity. We insert that formal series into  
\be 
E \equiv -D -D^\dagger + D^\dagger\,D \;.
\ee 
Finding a solution to the Ginsparg-Wilson equation  \cite{GiWi82} amounts to
solving for $E=0$. The product $D^\dagger\,D$  is again constructed
algebraically, leading to terms involving  Clifford algebra elements and paths
corresponding to ordered products  of gauge link variables. 
 
$E$ is an infinite series of terms that are linear and quadratic  in the
coupling coefficients of the Dirac operator multiplying the path terms.
Independent terms have to vanish and thus we obtain a system of coupled 
equations quadratic in the (unknown) coupling constants. We truncate the ansatz
for $D$ and the  series for $E$ at paths of length 4. Whereas the complete
series would lead to an  exact solution of the Ginsparg-Wilson equation, the
solution now will only be  approximate and it remains to be studied, how good
this approximation is. 

In solving the set of equations there is much freedom. In order to partially 
remedy the approximation we introduce two more conditions (and parameters) 
such that  the boundary condition for a massless Dirac operator 
\be\label{dfour}  \hat{D}(p) \equiv i \not \hspace{-1mm} p  +  {\cal O}(p^2)
\;,  \ee  is satisfied at a given value of the gauge coupling.  The solution we
study here has 19 terms and values of the coefficients are listed in the
appendix of \cite{Gaetal01a}. It is ultralocal with non-vanishing couplings
only for $|x-y|\leq \sqrt{5}$ decreasing exponentially in size. 

Exact solutions to the Ginsparg-Wilson equation ($E=0$)  have eigenvalues on a
unit circle with center 1 in the complex plane. It turned out that our
solution, defining a chirally improved Dirac operator, has eigenvalues, that
are very close to the unit circle in particular for the small eigenvalue
region, which is most relevant for the infrared behavior.  
 
{\bf Overlap operator:} 
Below we compare the properties of the sector with small eigenvalues  
(complex or real) of the introduced chirally improved operator with  
those of an overlap operator \cite{overlap} which has the form 
\begin{equation}\label{ovdef} 
D_{ov} = 1 - Z\quad \textrm{with}\quad Z\equiv\gamma_5\,\sign(H)\;, 
\end{equation} 
where $H$ is related to the hermitian Dirac operator. This is 
constructed from an arbitrary Dirac operator $D_0$ 
(e.g. the Wilson operator), 
\be 
H=\gamma_5 \,(s-D_0)\;, 
\ee 
and $s$ is a parameter which may be adjusted in order to minimize the 
probability for zero modes of $H$.   In case $D_0$ is already an overlap
operator one reproduces   $D_{ov}=D_0$ for $s=1$ because  
$\sign\left(\sign(H)\right)=\sign(H)$. 
 
The sign-function may be defined through the spectral representation 
\be 
\sign H = \sum_i\,\sign\lambda_i\,|i\rangle\langle i|\;, 
\ee 
($|i\rangle$ denote the eigenvectors) but in practical computations this
definition cannot be used for  realistic QCD Dirac operators, which for $L^4$
lattices have dimension  $\mathcal{O}(10^5-10^6)$  ($\sim n_{color}\cdot
n_{Dirac} \cdot L^4 $). Note  that in the  subsequent applications the operator
has to be applied many  times, since it may be entering a diagonalization or a 
conjugate gradient  inversion tool. One therefore relies on the relation 
\be  
\sign(H)=\frac{H}{\sqrt{H^2}} 
\ee 
and approximates the inverse square by some method. In our computations 
\cite{Gaetal01c} we follow the methods discussed in \cite{HeJaLu,Bu98}.  One 
approximates the inverse square root by a Chebychev polynomial, which has 
exponential convergence in $[\epsilon,1]$, where $\epsilon$ (and thus the
order  of the polynomial)  depends on the ratio of smallest to largest
eigenvalue of   $H^2$. Clenshaw's  recurrence formula provides further
numerical stability. 
 
Depending on the input Dirac operator $D_0$ the rate of convergence may be 
quite unfavorable. Although one may adjust $s$ it turns out that particular 
simple operators like the Wilson Dirac operator  eventually will give rise to 
several small eigenvalues of $H^2$. Technically, one then proceeds as follows: 
One determines the subspace of e.g.  the lowest 20 eigenmodes of $H^2$ and 
computes the inverse square separately for the subspace (using the spectral 
representation) and the reduced operator (with polynomial approximation). For 
this approach it is important to determine the subspace with high accuracy (we 
request 12 digits). 

In our study we need to diagonalize not only the hermitian matrix $H$ but, for 
the later analysis, also $D_{ov}$ and the chirally improved operator. All 
diagonalizations have been done with the Arnoldi method  \cite{arnoldi}. 
 
{\bf CPU-time comparison:} 
The central element of the diagonalization (and conjugate gradient methods) is 
the multiplication of the Dirac operator with a vector. For the consideration 
of the  computational effort we disregard the startup time for initialization 
of the operator. Comparing the CPU-time in units of the time used for the
Wilson Dirac operator we find that the chirally improved operator needs 24
units and the overlap operator (with Wilson input) typically needs  $300\pm
100$ units (always with 12 digits accuracy for the matrix norm  $|| H^2-1||$).
These are the results for lattice size $12^4$, but they are not  significantly
different for $8^4$ or $16^4$. 

As mentioned, whenever $(1-D_0)$ is already an overlap operator, one finds 
$D_{ov}=D_0$. We are interested in the case when $(1-D_0)$ is close to, but
not  quite an overlap operator. Such an operator might accelerate the
convergence in  the construction of the overlap operator, since  the
convergence of the polynomial approximation depends crucially  on the  smallest
eigenvalue of $H^2$. It is therefore of high practical  importance,  whether
$H$ will have eigenvalues close to zero. A Dirac operator   $D_0$ used as input
to the construction of $H$  may produce such small  eigenvalues. Physically
they are related to the appearance or disappearance of  an instanton. Different
Dirac operators will  differ in their sensitivity to  ``identify'' such
structures in the gauge configuration. 

When using the chirally improved operator -- in comparison to using the Wilson
Dirac operator -- as starting point we need typically a  factor 2 less
matrix-vector multiplications ($75\pm 25$ terms of the  Chebychev  series for
$12^4$ or $16^4$ lattices). One reason that this improvement factor  is not
higher may lie in the removal of the small eigenmode subspace, which is 
relatively more efficient for the ``bad'' Wilson operator used as input to the 
overlap construction. Another reason may be that the series approximation
forces all eigenvalues to the unit circle; since the chirally improved operator
has its large eigenvalues not as close to the  circle as the small ones, one
has to pay the price with longer polynomials.
 
\begin{figure}[tb] 
\epsfig{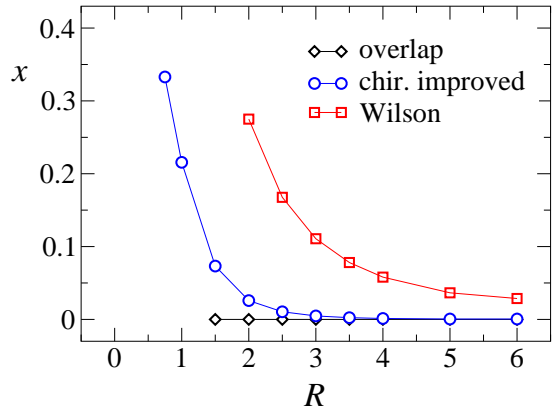}
\vspace*{-5mm}
\caption{Position $x$ of the real eigenvalue (the would-be zero mode) of 
different lattice Dirac operators as a function of the radius $R$ of the
underlying instanton configuration. 
\label{realdrift}} 
\end{figure} 
 
\section{Small eigenvalues and instantons} 
 
For an instanton in the continuum it is known that the Dirac operator has a
zero mode. On the lattice only an exact solution of the Ginsparg-Wilson
equation will display an eigenvalue which vanishes exactly. An approximate
Ginsparg-Wilson Dirac operator will then have a small  real eigenvalue.
Typically the size of the real part will increase as one shrinks the radius of
the underlying instanton. The rate at which the  real eigenvalue moves away
from the origin is a measure for the quality of the approximate solution of the
Ginsparg-Wilson equation.  A good approximation will keep the eigenvalue near
zero also for relatively small instantons. In order to implement this test of
the would-be zero modes we use instantons with several values of the radius $R$
discretized on the lattice following the method described in
\cite{FoLaSc85,Gaetal01c}.

Fig.~\ref{realdrift} shows the position $x$ of the real eigenvalue (the
would-be zero mode) of different Dirac operators as a function of the radius
$R$ of the underlying instanton. We show the results computed on $16^4$ 
lattices for the chirally improved operator (circles), the Wilson operator
(squares), and the overlap operator (diamonds). The latter, since it is an
exact solution of the Ginsparg-Wilson equation, has an exact zero mode. It is,
however, interesting to see that the overlap operator can trace the instanton
only down to a radius $R=1.5$. The Wilson operator has an eigenvalue which
very quickly departs from zero as $R$ is decreased. The chirally improved
operator keeps the real eigenvalue fairly close to 0 even for relatively small
instantons and only for $R \le 2$ the deviation becomes  significant.

For practical applications of the chirally improved Dirac operator, the
question is whether also in a gauge configuration with quantum fluctuations the
real eigenmodes lie close to the origin. In order to test this property, we
computed histograms for the probability distribution of the real eigenvalues.
The gauge fields were generated in a quenched simulation, using the tadpole
improved L\"uscher-Weisz action \cite{LuWeact}.  The histograms were computed
on $16^4$  lattices at values of $\beta_1 =$ 8.10, 8.30 and 8.45 using 200
configurations for each $\beta_1$. The corresponding values of the couplings
$\beta_2$ and $\beta_3$ of the gauge action can be found in \cite{Gaetal01b}.
There we obtained for the lattice spacing $a$ the values
0.127~fm, 0.107~fm, and 0.1~fm for $\beta_1=$ 8.10, 8.30, and 8.45,
respectively. 
  
\begin{figure}[tb] 
\epsfig{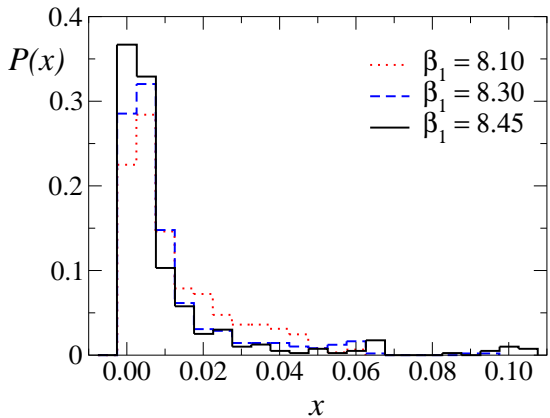}
\vspace*{-5mm}
\caption{Probability distribution of the position $x$ of the real eigenvalues
of the chirally improved Dirac operator. The data were computed on $16^4$
lattices at values of $\beta_1 = 8.10, 8.30$ and 8.45 using 200 configurations
for each $\beta_1$. 
\label{realdist}} 
\end{figure} 

From Fig.~\ref{realdist} it is obvious that the eigenvalues are concentrated
near the origin. The tail of the distribution decays very quickly and extends
almost exclusively towards values $x > 0$. Both these features are highly
welcome: The narrow width of the distribution lets one hope for only a very
small additive renormalization of the  quark mass. The fact that the
distribution does not extend to values $x<0$ implies furthermore that
exceptional configurations (which spoil the computation of the propagator) are
suppressed effectively.

\section{Localization properties of zero modes} 
 
We now analyze the eigenvectors with real eigenvalues, i.e.~the modes which
correspond to the zero modes in the continuum.  For a continuum instanton the
zero modes are known to be localized at the same position as the instanton. 
This property also holds for thermalized configurations on the lattice
\cite{chuetal}. The eigenmodes provide an efficient filter for the excitations
of the QCD vacuum as seen by the Dirac operator. 

To study the eigenvectors we define the scalar density $p_0(x)$. Let 
$\psi(x)_c$ be an eigenvector of the Dirac operator with $x$ denoting the
lattice point while $c$ is the color index. A gauge invariant density
$p_\sigma(x)$ is defined as
\begin{equation}
p_\sigma(x) =  \sum_c \psi(x)_c^* \,\Gamma_\sigma\,\psi(x)_c 
\;.
\label{scaldens}
\end{equation}
(Here $\Gamma_0=\mathbf{1}$, $\Gamma_5=\gamma_5$;  the sum over Dirac indices
is implied.)  In the last section we have seen that the continuum zero modes
correspond to  eigenvectors of the lattice Dirac operator with eigenvalue zero
(overlap operator) or small real eigenvalue (chirally improved operator, Wilson
operator). An interesting question is, to which extent the eigenvectors are
different for different lattice Dirac operators. 
 
\begin{figure}[tp] 
\begin{center}
\epsfig{file=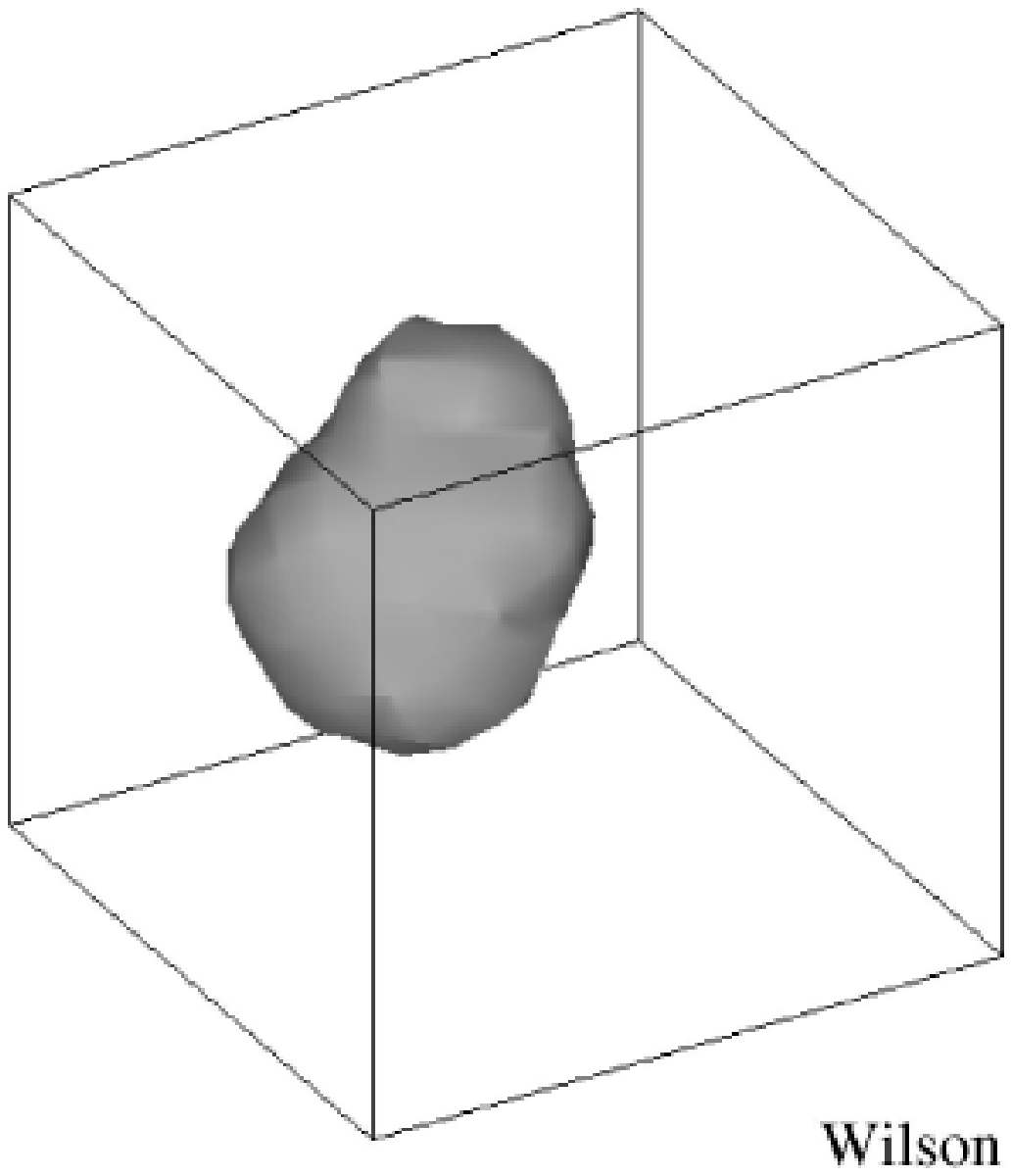,height=5.4cm,clip=}\\
\epsfig{file=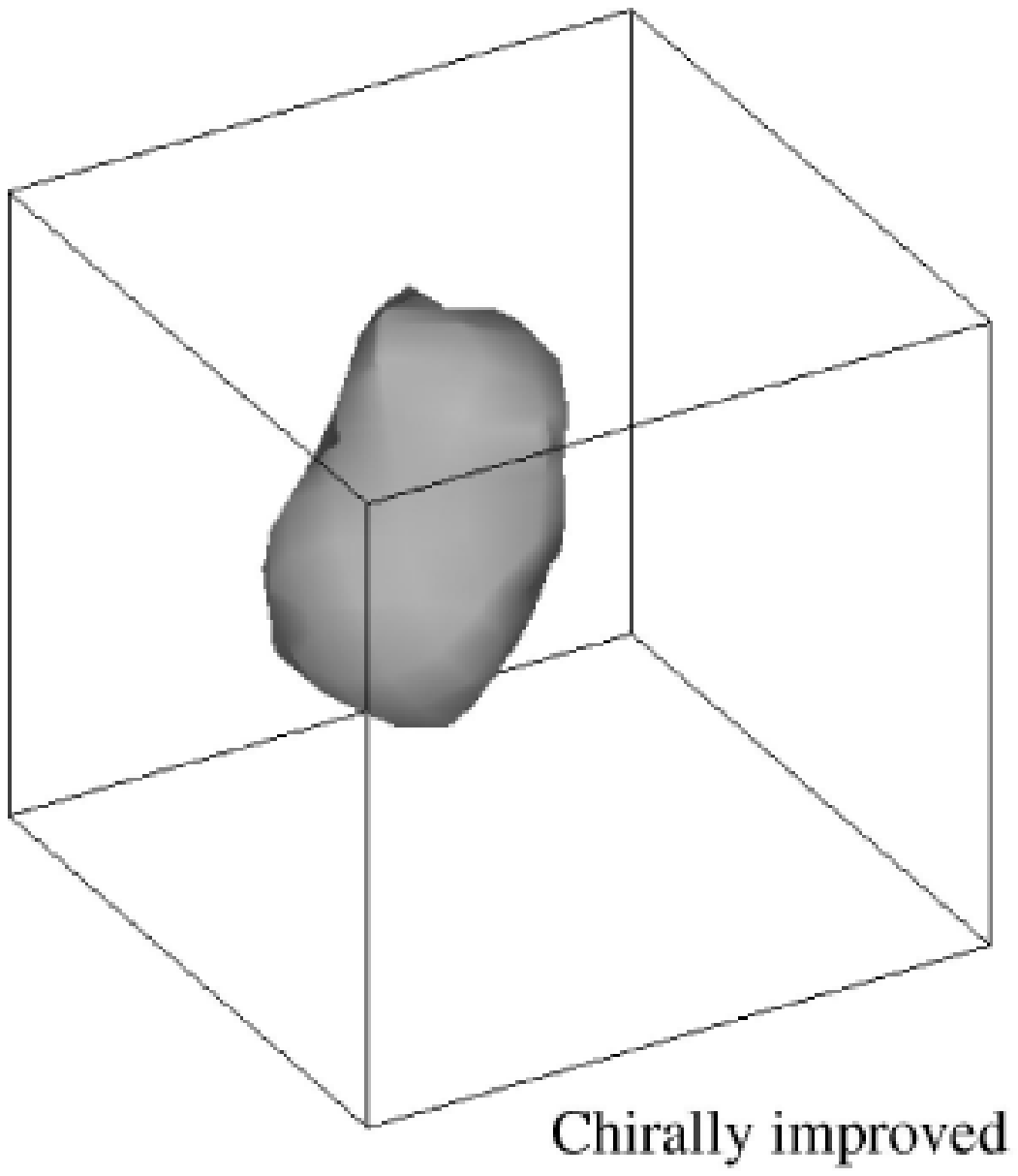,height=5.4cm,clip=}\\
\epsfig{file=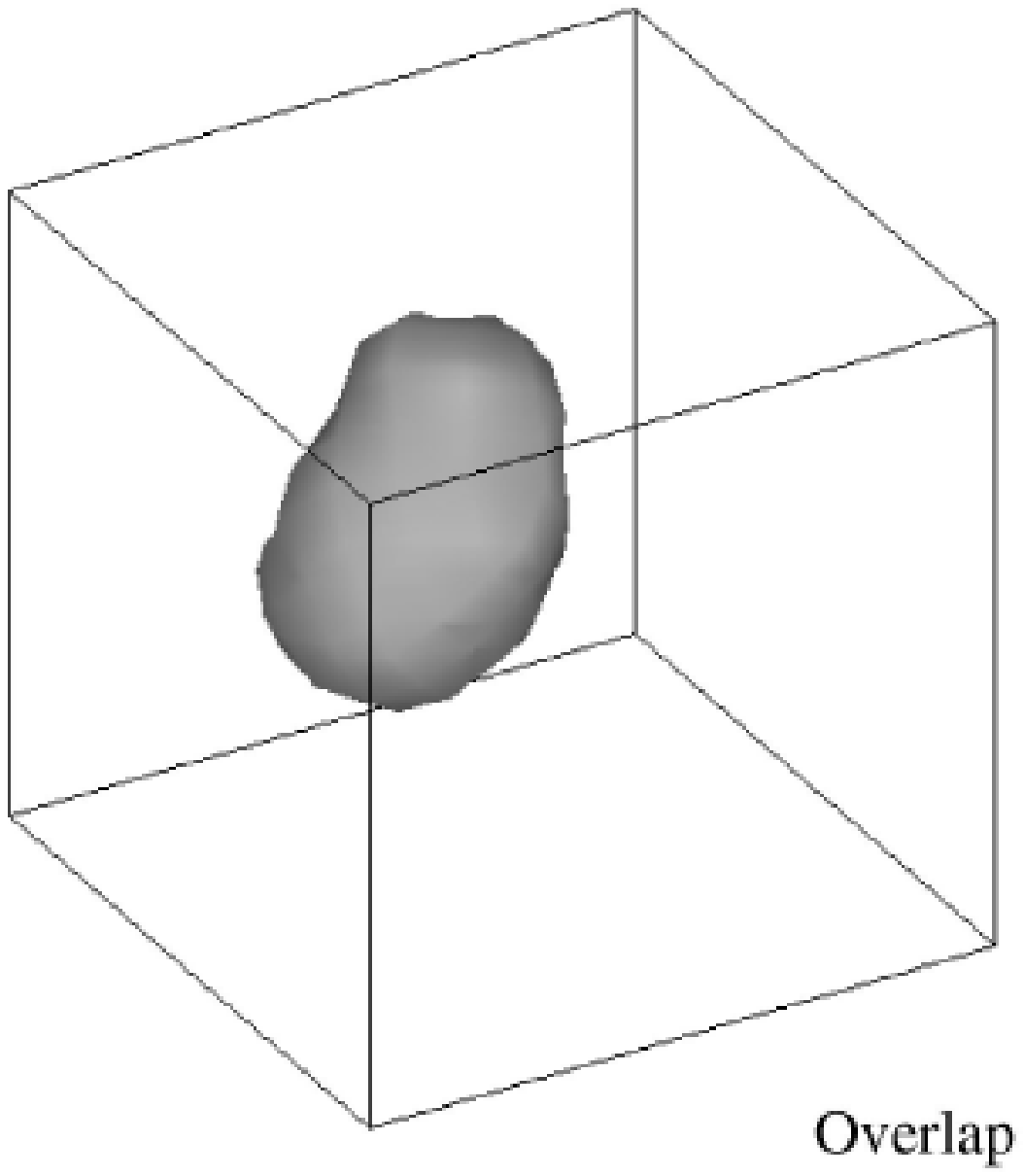,height=5.4cm,clip=}
\end{center}
\vspace*{-5mm}
\caption{Comparison of iso-surfaces (for a given time-slice) of the scalar density for
the zero-mode of different Dirac operators. We show the density for the 
Wilson operator (top), the chirally improved operator (middle), and the
overlap operator (bottom). 
\label{iso}} 
\end{figure} 

In Fig.~\ref{iso} we compare iso-surfaces of the scalar density for the
zero-mode of different Dirac operators on the same gauge configuration
($12^4$ lattice at $\beta_1 = 8.45$). All three operators see the same localized
instanton. Only the details vary slightly as can be seen by the different
fluctuations around the core. 

An interesting effect was observed for small instantons in \cite{Gaetal01c}
where we compared the localization of the zero mode for the smooth instantons
discussed above for different lattice Dirac operators. It was found that for
instantons with  radius smaller than 3 in lattice units, the overlap operator
has a zero mode which is less localized than in the case of a continuum instanton.
Since the other two operators (chirally improved, Wilson) do not display  such
a blowing up of zero modes we  attribute this effect to the rather big size of
the overlap operator. As opposed to the other two operators the overlap
operator is not ultra-local, i.e.~$|D_{ov}(x,y)|$ decays exponentially with
$|x-y|$ but does not vanish exactly. Objects smaller than the typical size of
the overlap operator thus can no longer be resolved. We estimate
\cite{Gaetal01c} that at typical lattice spacings of $a \sim$ 0.1~fm this
effect sets in at about 0.3~fm.

For further study we introduce the so-called inverse participation ratio, a
convenient measure of the localization  widely used in solid state physics. Due
to the normalization of the eigenmodes we have $\sum_x \, p_0(x) = 1$. The
inverse participation ratio is then defined as 
\begin{equation}
I \; = \; V \sum_x \, p_0(x)^2 \; ,
\label{ipr}
\end{equation}
where $V$ is the volume of the lattice. An alternative observable for
localization, based on the self correlation  of the scalar density $p_0(x)$ was
analyzed in \cite{degrandha}.

\begin{figure}[tb] 
\epsfig{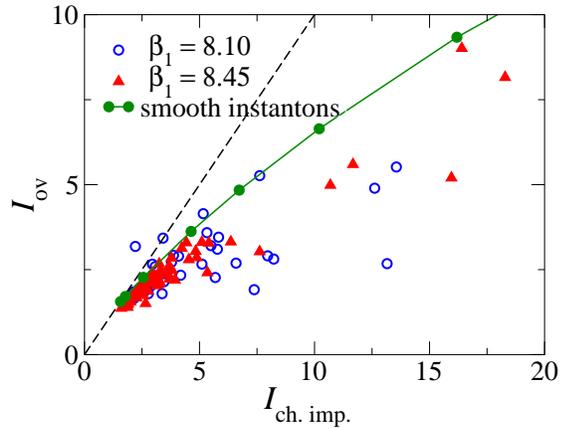}
\vspace*{-7mm}
\caption{Comparing the localization of the zero modes of the chirally improved
operator and the overlap operator. 
\label{iprcompare}} 
\end{figure} 

In Fig.~\ref{iprcompare} we show a scatter plot of the inverse participation
ratio for zero modes of the overlap operator $I_{ov}$ plotted as a function of
the inverse participation ratio of the corresponding modes of the chirally
improved operator $I_{ch.~imp.}$.  The data were computed on $12^4$ lattices
using 200 configurations at $\beta_1 = 8.10$ (circles), and $\beta_1 = 8.45$
(filled triangles). The full curve with filled circles is the result for the
smooth artificial instantons. The dashed line is a straight line with slope 1.
If both operators would see the same localization for the zero modes, all data
points would be located on this line. However, obviously this is not the case.
Except for a single data point, all data are below this line. This means that
the overlap operator systematically produces zero modes with smaller
localization when compared to the chirally  improved operator; this trend
becomes stronger for more localized states.

\section{Analysis of the near-zero modes} 
 
So far we have concentrated on analyzing the eigenvectors with real
eigenvalues, i.e.~the would-be zero modes. However, also the eigenvectors with
small but complex eigenvalues, the so-called near-zero modes, play an important
role in QCD phenomenology. In particular the eigenvalue density $\rho(\lambda)$
near at the origin  is related to the chiral condensate through the
Banks-Casher formula \cite{BaCa80},
\begin{equation}
\langle \overline{\psi} \psi \rangle \; = \; - \pi
\lim_{\lambda\to 0}\lim_{V\to\infty} \rho(\lambda)V^{-1} \; .
\label{bcformula}
\end{equation}

\begin{figure}[tb] 
\vspace*{-1mm}
\epsfig{file=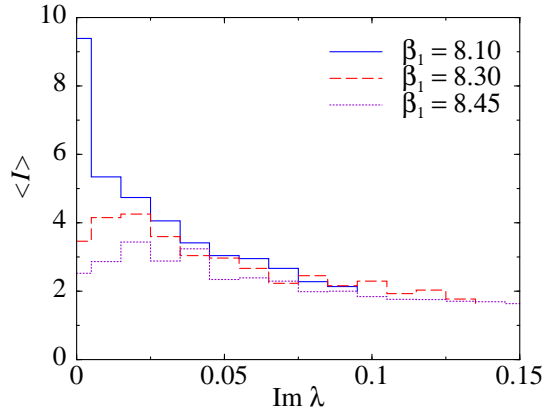,width=73mm,clip}
\vspace*{-4mm}
\caption{Histograms of the inverse participation ratio for 
eigenvectors as a function of the imaginary part of the corresponding
eigenvalue. 
\label{iprvslambda}} 
\end{figure} 

According to the instanton liquid picture of chiral symmetry breaking
\cite{instpheno}, the near-zero  modes come from weakly interacting instantons
and anti-instantons. The corresponding eigenvectors are expected to be still
quite localized, since  they correspond to interacting ``lumps'' which still 
resemble instantons or anti-instantons. These predictions may be
tested in an ab-initio calculation on the lattice. Again we use the inverse
participation ratio defined in Eq.~(\ref{ipr}) to study the localization of
these modes. An interesting question is how the localization of eigenvectors
varies as a function of the eigenvalue. 

In Fig.~\ref{iprvslambda} we show a histogram of the inverse participation
ratio   for eigenvectors as a function of the imaginary part of the
corresponding eigenvalues. The data were computed on $16^4$ lattices at values
of $\beta_1 = 8.10, 8.30$ and 8.45 using 200 configurations for each $\beta_1$.
It is obvious that the most localized states have eigenvalues near the origin.
Such large values are expected for weakly  interacting instantons and
anti-instantons. As $\textrm{Im\,}\lambda$ is increased, the localization
decreases. The corresponding modes do no longer feel distinguished topological
lumps but become dominated by quantum fluctuations.

\begin{figure}[tb] 
\epsfig{file=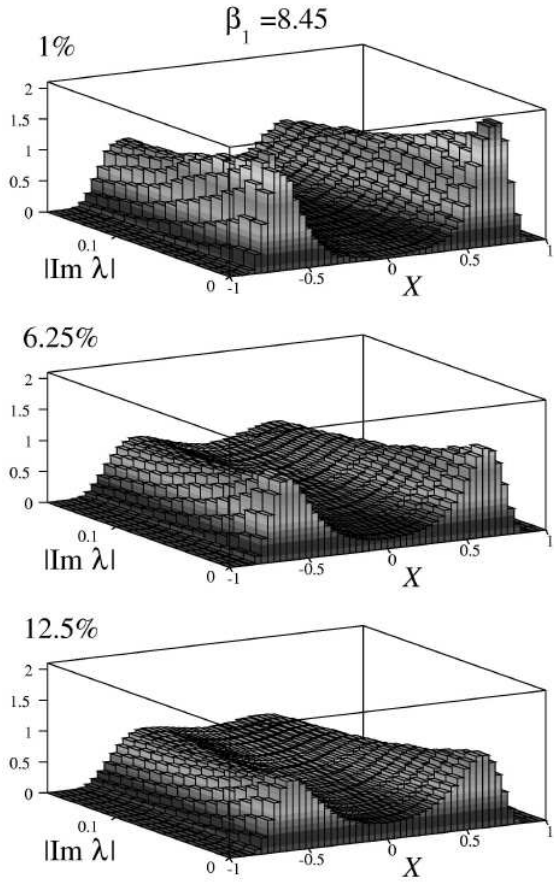,width=74mm,clip}
\vspace*{-5mm}
\caption{Local chirality of eigenvectors as a function of 
$\textrm{Im\,}\lambda$ of the corresponding eigenvalues. We use cuts of 1\% (top)
6.25\% (middle) and 12.5 \% (bottom) for the number of supporting
points.
\label{tub}} 
\end{figure} 

An interesting observable was suggested by Horvath et al.~in \cite{locchir}.
Their local chirality variable was subsequently studied by several other groups
in \cite{locchir2,Gaetal01a,Gaetal01b}. The underlying idea is to test whether
the near-zero modes are locally chiral, i.e. left-handed for lattice points
near the instanton peak of the density and right-handed near the anti-instanton
peak. The construction starts with defining left- and  right-handed densities
$p_\pm(x)$,
\begin{equation}
p_\pm(x) =\frac{1}{2}\left( p_0(x)\pm p_5(x)\right)\;,
\end{equation}
i.e. the density (\ref{scaldens}) projected on positive and negative chirality.
A mode dominated by instantons and anti-instantons is expected to have 
$p_+(x) > 0$ and $p_-(x) = 0$ for lattice points $x$ near the instanton peak
and  vice versa for $x$ near an anti-instanton peak. Thus the  ratio
$p_+(x)/p_-(x)$ should be large for $x$ close to an instanton peak and small
for all $x$ close to an anti-instanton peak. In a final step Horvath et~al. map
the positive real values of this ratio onto the interval  between $-1$ and $+1$
using the transformation
\begin{equation}
X(x) \; = \; \frac{4}{\pi} \; 
\mbox{arctan} \, \left( \sqrt{\frac{ p_+(x) }{ p_-(x) }} \right) \; - 
\; 1 \; .
\end{equation}
In order to suppress the quantum fluctuations one evaluates $X(x)$ only for $x$
near instanton or anti-instanton peaks, i.e.~for $x$ near the maxima of the
scalar density $p_0(x)$. We computed our results using cutoffs of 1\%,  6.25\%
and 12.5\% for the number of lattice points $x$ (ordered according to $p_0(x)$)
and averaged over $X(x)$.  On the gauge field ensemble the local  chirality
variable is expected to show a double peak structure only for instanton
dominance.

In Fig.~\ref{tub} we show the local chirality of the near-zero modes as a
function of the imaginary part of their eigenvalues.  The data were computed
from 200 configurations on $16^4$ lattices at $\beta_1 = 8.45$. When using a
cut of 1\% on the number of supporting lattice points (top plot) we find the
best signal while for cuts of 6.25\% (middle plot) and 12.5\% (bottom plot) the
signal becomes weaker due to quantum fluctuations. Furthermore it is obvious
for all three plots that the strongest signal for local chirality is found for
the eigenvectors with eigenvalues closest to the origin. For larger values of
$|\textrm{Im\,}\lambda|$  the signal of local chirality for the corresponding
eigenvectors weakens and the mode becomes dominated by quantum fluctuations. 

\begin{figure}[tb] 
\epsfig{file=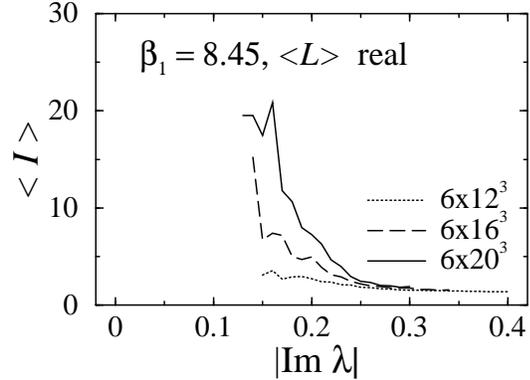,width=70mm,clip}
\vspace*{-5mm}
\caption{The inverse participation ratio as a  function of the imaginary part
of the corresponding eigenvalue. These data were computed at $\beta_1 = 8.45$
where  (with time extent 6) the theory is in the deconfined phase. In the
subsample used here the expectation value of the Polyakov loop $\langle L
\rangle$ is close to real.
\label{ivslambdadec}} 
\end{figure} 

\section{The near-zero modes at finite temperature} 

Let us now discuss what happens to instantons at the QCD phase transition. At
the critical temperature $T_c$ the theory enters the  deconfined phase and
chiral symmetry is restored. The spectrum develops a gap  (see
e.g.~\cite{Gaetal01a} for a plot of typical spectra) and the density of
eigenvalues is zero near the origin. According to the Banks-Casher formula
(\ref{bcformula}) this gap in the spectral density amounts to a vanishing
chiral condensate.

The topological charge does not vanish abruptly at $T_c$ and topological
objects are observed also for $T  >  T_c$. Thus the liquid of weakly
interacting instantons responsible for chiral symmetry breaking at $T < T_c$
must undergo some change in its structure as the temperature is increased above
$T_c$. It may be expected that above $T_c$  instead of interacting weakly,
instantons and anti-instantons form tightly  bound molecules \cite{instpheno}.
This prediction from instanton models can be tested on the lattice (compare
\cite{regensburg} for a similar study using the staggered Dirac operator). 

In Fig.~\ref{ivslambdadec} we show the inverse participation ratio  as a
function of the imaginary part of the corresponding eigenvalue, as in
Fig.~\ref{iprvslambda}, but now for finite temperature. In particular we show
the results for configurations where the Polyakov loop $\langle L \rangle$ is
in its real branch, which is the domain of the Polyakov loop value for the
full, unquenched theory. For the equivalent plot with  complex Polyakov loop
see \cite{Gaetal01a}. Due to the spectral gap in the deconfined phase, there 
are no near-zero modes and the graph has support for  $|\textrm{Im\,}\lambda| 
>  0.15$. One finds that now the most localized states  are near the edge of
the spectrum. Localization decreases quickly as  $|\textrm{Im\,}\lambda|$ is
increased and the mode becomes a bulk mode dominated by quantum fluctuations.
The inverse participation ratio, however, provides only information on the
localization of the modes and does not test for other properties such as local
chirality. Thus it makes sense to repeat  last section's  study of Horvath et
al.'s local chirality variable for  $T > T_c$. 

\begin{figure}[tb] 
\epsfig{file=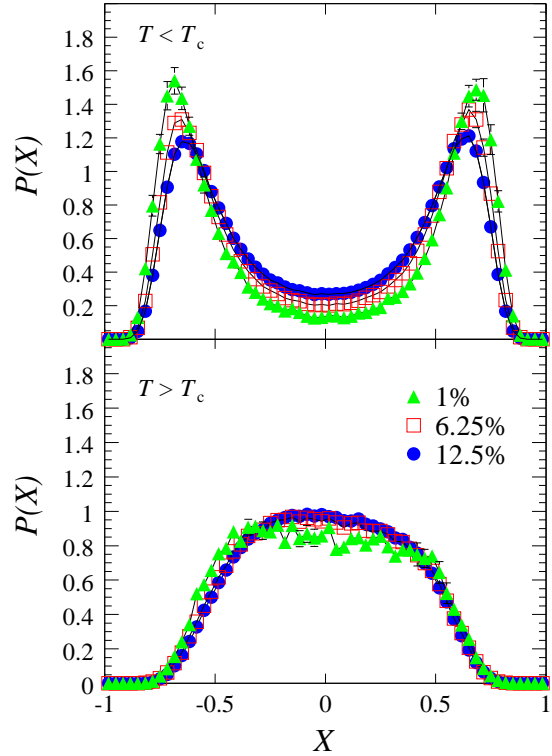,width=73mm,clip}
\vspace*{-5mm}
\caption{Comparison of the local chirality of the (complex)
near-zero modes for
$T < T_c$ (top) and $T > T_c$ (bottom). 
\label{locchirtemp}} 
\end{figure} 

In Fig.~\ref{locchirtemp} we compare the local chirality of the near-zero modes
for an ensemble at $T <  T_c$ (top plot, $6\times20^3$ lattices, 400
configurations at $\beta_1 = 8.10$) with an ensemble at $T  >  T_c$ (bottom
plot,  $6\times20^3$ lattices, 147 configurations with real branch Polyakov
loop at $\beta_1 = 8.45$).  The plots were computed using only the
eigenvectors  with  $|\textrm{Im\,}\lambda|$ less than some cut corresponding
to roughly the  4 eigenvalues closest to the origin, respectively the edge of
the spectrum. Zero modes were left out. For $T <  T_c$ we find a very
pronounced double peak structure, i.e.~a  strong signal for local chirality. At
$T > T_c$ the double peak structure is gone and we do not observe local
chirality in the eigenmodes. It is obvious that the weakly interacting
instantons and anti-instantons undergo a drastic change as $T$ increases above
$T_c$. The resulting states are still localized  (see Fig.~\ref{ivslambdadec})
but the local chirality present for weakly interacting topological excitations
has vanished, probably due to the formation of tightly bound molecules.

\vspace*{3mm}

{\bf Acknowledgements:} We thank the Leibniz Rechenzentrum in Munich for
computer time on the Hitachi SR8000 and their operating team for support and
training. C.~Gattringer and C.B.~Lang thank the DOE's Institute of Nuclear
Theory at the University of Washington for its hospitality and the DOE for
partial support during the completion of this work. We have profited
considerably  from discussion with Pierre van Baal, Tom DeGrand, Stefan
D\"urr,  Peter Hasenfratz, Ivan Hip and Ferenc Niedermayer.

\end{document}